\documentclass[11pt]{article}
\usepackage[dvips]{graphicx}
\usepackage{amssymb}
\usepackage{amsmath}
\usepackage{amssymb}
\usepackage{overcite}
\usepackage{epsf}
\input{epsf}
\begin{document}
\begin{titlepage}
\title{\bf Thermodynamic Limit in Statistical Physics\thanks{Int. J. Mod. Phys. B, 
Vol.28, (2014) p.1430004 (28 pages);   DOI: 10.1142/S0217979214300047}}  
\author{A. L. Kuzemsky 
\\
{\it Bogoliubov Laboratory of Theoretical Physics,} \\
{\it  Joint Institute for Nuclear Research,}\\
{\it 141980 Dubna, Moscow Region, Russia.}\\
{\it E-mail:  kuzemsky@theor.jinr.ru} \\
{\it http://theor.jinr.ru/\symbol{126}kuzemsky}}
\date{}
\maketitle
\begin{abstract}
The thermodynamic limit in statistical thermodynamics of many-particle systems is an important
but often overlooked issue in the various applied studies of condensed matter physics. To settle this issue,
we review tersely  the past and present disposition of thermodynamic limiting procedure in the structure
of the contemporary statistical mechanics and our current understanding of this problem.
We pick out the ingenious approach by  N. N. Bogoliubov, who developed a general
formalism  for establishing of the limiting distribution functions in the form of formal series in powers
of the density. In that study he outlined the method of justification of the 
thermodynamic limit when he derived  the generalized Boltzmann equations.
To enrich and to weave our discussion, we take this opportunity to give a brief survey of the closely related 
problems, such as  the equipartition of energy and the equivalence and nonequivalence  of statistical ensembles.  
The validity of the equipartition of energy permits one to decide what are the boundaries of applicability of statistical
mechanics.
The major aim of this work is to provide a  better qualitative understanding of the physical significance of the
thermodynamic limit in modern statistical physics of the infinite and "small" many-particle systems.
\vspace{1cm}

%
%
%
\vspace{0.5cm}

%
%
\end{abstract}
\end{titlepage}
\newpage
\tableofcontents
\newpage
%
%
%
%
%
%
%
%
\section{Introduction}\label{}
%
%
%
Equilibrium statistical mechanics~\cite{gibbs,jhj25,rice30,tolm,pmor,petqsm95,bwid02,dorl,dpet} is a well explored and relatively 
well established subject, in spite of some unsettled foundational issues. However, it was claimed about decade ago in an
authoritative scientific journal~\cite{vran98} that
\begin{quote}
\label{ } \em The fact that classical  equilibrium statistical mechanics works is deeply puzzling.
\end{quote}
The purpose of the present paper will be to elucidate certain aspects of the reconciliation between the 
statistical mechanics and mechanics, i.e. dynamical systems~\cite{carat07,koz08}, to emphasize and address
a few important reasons for such a workability. Our central interest here will be the thermodynamic limit, 
equipartition of energy and equivalence and nonequivalence of ensembles.\\
In classical equilibrium statistical thermodynamics 
one deals with equilibrium states of a system. It is assumed that each of those states corresponds to a set of 
indistinguishable microstates, because the temperature, the pressure, and all other the so-called thermodynamic variables
have the same value for each microstate of the set. Quantities, such as pressure and temperature are termed the state variables, 
which characterize the system in the state of statistical equilibrium. The thermodynamic limit is reached when the 
number of particles (atoms or molecules) in a system tends to infinity.\\
Hence, in statistical physics, the  \emph{thermodynamic limit}  denotes the limiting behaviour of a physical system that consists of 
many particles   (or components) as the volume $V$ and the number $N$ of particles tends to infinity. 
Simultaneously, the density ratio $N/V \sim  n $ approaches a constant value. 
Many characteristic properties of macroscopic physical systems only appear in this limit, namely phase transitions, 
universality classes and other critical phenomena.\\ 
It is worth noting that  the problem of the thermodynamic limit at the earlier  stage of statistical mechanics 
was  hid behind many technicalities of the new   discipline~\cite{gibbs,jhj25,rice30,tolm,pmor}. Some part of
modern textbooks do the same. Contrary to this, other   modern textbooks (see, e.g., Refs.~\cite{petqsm95,bwid02,dorl})
discuss the thermodynamic limit carefully and with eminently suitable manner. For example, the textbook 
by  Widom~\cite{bwid02} mention the thermodynamic limit by nine times and book by Dorlas~\cite{dorl}
devotes to this question the special chapter. It is remarkable, that  the first time when the notion of the 
thermodynamic limit appears in the Widom's book~\cite{bwid02}, is that when he derives the celebrated 
Rayleigh-Jeans law~\cite{jhj25}. These authors~\cite{petqsm95,bwid02,dorl} demonstrated  explicitly the essential role of 
the thermodynamic limit (which has already been presented in an implicit form in Jeans book~\cite{jhj25}) for the 
consistent derivation of that law and other  important issues of statistical mechanics.\\
A significant step in the rigorous treatment of the thermodynamic limit was made by 
N. N. Bogoliubov, who developed a general
formalism  for establishing of the limiting distribution functions in the form of formal series in powers
of the density.
In his famous monograph~\cite{bogol46,bog62}, Bogoliubov outlined the method of justification of the thermodynamic limit 
and derived the generalized Boltzmaun equations from his formalism (see also Refs.~\cite{dpet,dpet09,gera97}). For this purpose, he introduced the concept of stages of
the evolution-chaotic, kinetic, and hydrodynamic  and the notion of the time scales, namely, interaction time, free path
time, and time of macroscopic relaxation, which characterize these stages, respectively. At the chaotic stage,
the particles synchronize, and the system passes to local equilibrium. He showed then, that at the kinetic stage, 
all distribution functions begin to depend on time via the one-particle function. Finally, at the hydrodynamic stage, 
the distribution functions depend on time via macroscopic variables, and the system approaches equilibrium.
Bogoliubov also introduced the important clustering principle.
Furthermore, these distribution functions, which  are equal to the product of functions, one of which depends only on momenta being
indeed the Maxwell distribution, and the second one depends only on coordinates.  Bogoliubov conjectured that 
it is often convenient to separate the dependence on momenta and consider distribution functions, 
which will depend only on coordinates passing then to the thermodynamic limit. Thus, on the basis of his equations 
for distribution functions   and the cluster property, the Boltzmann equation
was first obtained without employing the molecular chaos hypothesis.\\
Indeed, let us consider~\cite{dpet,dpet09} the state of a finite system, which consists of $N$ particles distributed
with density $1/V$ in a region $\Lambda$ with volume $V$, $|\Lambda| = V$. The system is described by a probability 
distribution function $F_{N, \Lambda}(t, x_{1},x_{2} \ldots x_{N})$ given on the phase space $x = (p, q)$, 
where $p$ is a momentum, and $q$ is a coordinate. This function is defined as the solution of the corresponding 
Liouville equation, which satisfies   certain  initial conditions, described in Refs.~\cite{bogol46,bog62}
The interaction potential $\Phi (q_{i} - q_{j})$ was supposed to be the pairwise.\\
The average value of an observable $A_{N}(t, x_{1},x_{2} \ldots x_{N})$, where $A_{N}$ is a real symmetric function, 
is given by the formula
\begin{equation}\label{1.1}
 \langle A_{N} (t) \rangle = \int A_{N}(t, x_{1},x_{2} \ldots x_{N}) F_{N, \Lambda}(t, x_{1},x_{2} \ldots x_{N})
d x_{1} d x_{2} \ldots d x_{N}.
\end{equation}
The state of an infinite system is obtained as a result of the thermodynamic limit procedure under which the
number of particles $N$ and the volume $V$ of the region $\Lambda$ tend to infinity while the density remains constant:
$$N \rightarrow \infty, \quad  V \rightarrow \infty, \quad N / V = n. $$
A rigorous proof of the existence of the thermodynamic limit appeared to be a very difficult problem~\cite{dpet,dpet09}. To clarify
the nature of the difficulties, it is worth noting that the distribution functions $F_{N, \Lambda}$ are equal 
to the ratio of the variables which diverge as $N^{N}$ in the thermodynamic limit. Thus it was necessarily to prove 
that these divergences compensate each other and that
the limiting distribution functions will be really well defined as a mathematical object. 
The main formulas obtained for equilibrium
distribution functions correspond to Gibbs  results, however the problem of justification of the thermodynamic limit
procedure remained unsolved for about 50 years because of the difficulties described above. Only in 1949 did
Bogoliubov propose the solution of this problem~\cite{bog49}. He reduced it to the functional-analysis problem of proving the
existence of solutions to certain operator equations and investigating their limiting properties. This program was
realized on the basis of equations for distribution functions~\cite{bpk69,petri89,petr09}.\\
In the present topical review a brief survey of some important questions concerning the thermodynamic limit and related
problems   will be carried out.  
Our main intention is to sketch here the physical results rather than a mathematical formalism.
Hence, we will stay away from technicalities and will concentrate on the essence of 
the problems from the physical viewpoint.
%
%
%
%
\section{Interrelation of Statistical Mechanics and Thermodynamics}\label{}
%
%
Before considering of special questions a very brief summary of the interrelation of statistical mechanics and 
thermodynamics will be instructive.\\
The aim of statistical mechanics is to give a consistent
formalism for a microscopic description of   macroscopic behavior of matter 
in bulk~\cite{gibbs,hill56,hill60,es89,donald00,gallsm,bb,zub74}. 
The central problem in the statistical physics of matter is that of accounting for the observed equilibrium and nonequilibrium
properties of fluids and solids from a specification of the component molecular species, knowledge of how the constituent molecules interact,
and the nature of their surrounding.  The methods of equilibrium and nonequilibrium statistical mechanics have been fruitfully applied to a large variety of phenomena and 
materials. From the other side, during the last decades there was a substantial progress in  mathematical foundations of statistical 
mechanics~\cite{dorl,dpet,khin,leb68,minl,minl2,dobr94,dobr,open,dru99,drue,gala04,kuz07} and in studies of ergodic theory and 
theory of  dynamical systems~\cite{gala04,kuz07}.\\
It is known that thermodynamic properties of many-particle systems are the physical characteristics that are selected
for a description of systems on a macroscopic scale~\cite{tolm,donald00,hill56,hill60,hon3}. Classical thermodynamics~\cite{hon3}
considers the systems (i.e. a region of the space set apart from the remainder part for special study) which are in
an equilibrium state. Thermodynamic equilibrium is a state of the system where, as a necessary condition, none of its
properties changes measurably over a period of time exceedingly long compared to any possibly observations on the system. 
Classical equilibrium thermodynamics deals with thermal equilibrium
states of a system, which are completely specified by the small set of variables, e.g., by the volume $V$, internal energy $E$
and the mole numbers $N_{i}$ of its chemical components.\\
In classical statistical mechanics  one considers the number of particles $N$ which  is very large  
(typically of order $10^{23}$), enclosed in a finite but macroscopically large volume $V$. A reduced description
requires much smaller number of variable to operate with.
Thus,  construction of  statistical ensembles~\cite{gibbs,tolm,hill56,hill60} in the case of statistical equilibrium is based on the appropriate
choice of relevant integrals of motion on which the distribution function can depend.\\
The statistical ensemble is specified by the distribution function $f(p,q,t)$, which has the meaning of 
the \emph{probability density} of the distribution of systems in phase space $(p,q)$. More precisely, the distribution function
should be defined in such a way that a quantity 
$$ d w = f(p,q,t) d p d q$$ can be considered as the \emph{probability} of finding the system at time $t$ in the
element of phase space $d p d q$ close to the point $(p,q)$.\\
The thermodynamic variables with a \emph{mechanical} origin such as the internal energy $E,$ the volume $V,$ and the number of 
particles $N,$ are given well-defined values or \textbf{averages} of the mechanical quantities over the ensemble under 
consideration~\cite{gibbs,tolm,hill56,hill60}. On the contrary, thermodynamic variables such as the entropy $S,$ the temperature $T,$ and the chemical potential
$\mu$ do not have a mechanical nature. Those values are usually introduced by identifying terms in the fundamental
differential relation~\cite{tolm,donald00,hill56,hill60,hon3} for the energy  $E$
\begin{equation}\label{2.1}
  d E = T d S - P d V + \mu d N.
\end{equation}
Here $P$ is the pressure, one of the thermodynamic intensive variables, $T$ is the temperature and $\mu$ is the
chemical potential. Intensive (extensive) variables are the variables
whose value is independent of (depends on) the size and the quantity of matter within the region which is 
being studied~\cite{hon3}.\\
Contrary to this, the subject of statistical mechanics  aims to base  the statistical approach on the  microscopic models of
matter; it deals with those properties of many-particle systems which are describable 
in average~\cite{gibbs,tolm,hill56,hill60}. \\
There are mainly three methods used in equilibrium statistical mechanics,
namely, the Boltzmann method~\cite{gallsm,boltz,boltz1} of identifying the equilibrium state
with the most probable one; the Gibbs ensemble method~\cite{gibbs,tolm} of \emph{postulating}
a canonical distribution, and the Darwin-Fowler method~\cite{huang}  of identifying
the equilibrium state  with the average state. Schr\"{o}dinger~\cite{es89} termed the last approach by 
the \emph{method of mean values}.
It should be noted that the Darwin-Fowler method in statistical mechanics is a powerful method which allows in a straightforward 
way the evaluation of statistical parameters and distributions in terms of relatively simple contour integrals of certain generating 
functions in the complex plane.\\
As a result of the Gibbs ensemble method, the entropy $S$ can be expressed~\cite{tolm,hill56,hill60,es89,huang,etj65,grand08} in the form  of an average for all the ensembles, namely,
\begin{equation}\label{2.2}
  S (N, V, E) =  - k_{B} \sum_{i}p_{i} \ln p_{i} = - k_{B} \Omega \Bigl ( \frac{1}{\Omega} \ln \frac{1}{\Omega} \Bigr ) = 
 k_{B}  \ln \Omega (N, V, E),
\end{equation}
where the summation over $i$ denotes a general   summation over all states of the system and $p_{i}$  is the probability of
observing state $i$ in the given ensemble and $k_{B}$ is the Boltzmann constant. This relation links entropy $S$ and
probability $p_{i}$.
For thorough mathematical discussion and precise definition of Gibbs entropy see Ref.~\cite{mack06}\\ 
Boltzmann has used~\cite{hill56,hill60,es89,donald00,gallsm} a logarithmic relation in the following form
\begin{equation}\label{2.3}
  S =   k_{B} \ln \Omega.
\end{equation}
Here $\Omega$ is the probability of  a macroscopic state $E$  and 
$k_{B} = R/N_{A} = 1.3806 \cdot 10^{-23} \, J K^{-1}$ is the ratio of the molar gas 
constant $R$ to the Avogadro constant $N_{A}$ and has the dimension of entropy. It was termed the Boltzmann
constant; in essence this constant relates macroscopic and microscopic physics. Indeed, the ideal gas equations
are $P V = N k_{B} T$ and $U = x N k_{B} T$, where $x = 3/2 $ for a monoatomic gas, $x = 5/2 $ for a diatomic gas,
and $x = 6/2 $ for a polyatomic gas. Here $U$ is the internal energy of the gas.\\
Note that original Boltzmann expression $S =   k  \ln W$  defines the entropy $S$, a macroscopic
quantity, in terms of the multiplicity $W$ of the microscopic degrees of freedom of a system.
Since entropy is an additive quantity and 
probability is a multiplicative one, this relationship looks very natural 
(see Refs.~\cite{etj65,grand08,mack06,etj1,etj2,etj3,etj4,akatz,grand87,grand88,vank2,athe} for detailed discussion). 
It is easy to see that any monotonic function of $W$ will have a maximum where W has a maximum.
In particular, states that maximize $W$ also maximize the entropy, $S =   k  \ln W$.\\
The assumption of complete statistics~\cite{etj4} implies that all states regarding
the system is countable and known completely by us so that we have full knowledge of the interactions taking place in
the system of interest, thereby implying the ordinary normalization condition  $\sum_{i}p_{i} = 1$.  
An alternative procedure for the development of the statistical mechanical ensemble theory is to introduce the Gibbs
entropy postulate which states that for a general ensemble the entropy is given by Eq.(\ref{2.2}). Thus the postulate of 
equal probabilities in the microcanonical ensemble and  the Gibbs entropy postulate can be considered as a convenient 
starting point for the development of the statistical mechanical ensemble theory in a standard approach. It is should be said that this course
of development is workable  when the Boltzmann $H$-theorem was first established~\cite{gallsm,boltz,boltz1,uff09}.\\
After postulating  the entropy by means of Eq.(\ref{2.2}), the thermodynamic equilibrium ensembles are determined by the
following criterion for equilibrium:
\begin{equation}\label{2.4}
 \left ( \delta S \right )_{E,V,N} =  0.
\end{equation}
This variational scheme is used for each ensemble (microcanonical, canonical and grand canonical) with different constraints
for each ensemble~\cite{tolm,hill56,hill60,es89,donald00}. In addition, this procedure introduces Lagrange multipliers which, in turn, must be identified with
thermodynamic intensive variables ($T, P$) using by Eq.(\ref{2.4}). From the other hand, the procedure of introducing
Lagrange multipliers and the task of identifying them with the thermodynamic intensive properties can be clarified
by invoking a more general criterion for thermodynamic equilibrium~\cite{hon3,athe}. \\
It is worth noting that a close relationship exists between the concepts of 
entropy and probability~\cite{etj65,grand08,etj1,etj2,etj3,etj4,akatz,grand87,grand88}, 
the most famous of which is associated with the name of Boltzmann~\cite{gallsm,boltz,boltz1}.
Thus entropy and probability are intrinsically related~\cite{etj4}.  
It can showed that the concavity property of the entropy~\cite{lgal68,ascot69,galg70,ptl78,prest03} is directly related
to a given probability distribution function for an ideal gas in which binary collisions dominate. 
Concavity is directly related to the logarithm of a probability distribution.
It is interesting that by relating the entropy directly to a probability distribution function, one can show that
a non-equilibrium version of the entropy function may be deduced.\\ 
The very important statement of the Gibbsian statistical mechanics is the so-called \emph{Gibbs' theorem} on canonical
distribution. The theorem states that a small part of a microcanonical ensemble of systems with many degrees of freedom
is distributed canonically, i.e.  according to the law 
\begin{equation}\label{2.5}
  f(p,q) =    Q^{-1} (\theta, V, N) \exp \Bigl ( - \frac{H(p,q)}{\theta} \Bigr ).
\end{equation}
Here $Q  (\theta, V, N)$ is the partition function and $\theta$ is the modulus of the canonical  distribution which
corresponds to the temperature in the phenomenological thermodynamics~\cite{hon3}.  Thus the partition function $Q  (\theta, V, N)$
is an essential characteristic of the canonical Gibbs ensemble, which determines the thermodynamic properties of the
system. The partition function satisfies to the normalization condition
\begin{equation}\label{2.6}
  Q (\theta, V, N) =   \int  \exp \Bigl ( - \frac{H(p,q)}{\theta} \Bigr ) d \Gamma; \, d \Gamma = \frac{dp dq}{N! \hbar^{3N}}.
\end{equation}
As it was mentioned earlier, a statistical ensemble of systems with a specified number of degrees of freedom  $N$ and volume $V$
in contact with a thermal bath (which is a non-trivial notion~\cite{koz08,galla09}) is called a canonical Gibbs ensemble. 
Note, that the Gibbs' postulate~\cite{gibbs,tolm,donald00,hill56,dpet} states that the canonical
equilibrium distribution, of all the normalized distributions having the same mean energy, is the one 
with \emph{maximum entropy}~\cite{etj65,grand08,etj1,etj2,etj3,etj4,akatz,grand87,grand88,athe}.
In addition, the Gibbs' postulate rests on two assumptions. First, the stationary equilibrium distribution, being canonical, is
of exponential form. Second, Gibbs assumed that all the compared distributions have the same mean energy values. Thus the use of
a more general condition Eq.(\ref{2.6}) instead of Eq.(\ref{2.4}) as a criterion for thermodynamic equilibrium permitted
us treat the thermodynamic temperature $T$ directly in the framework of the statistical mechanical formulation.\\
Before closing this section, it will be informative to remind the important remark by Hugenholtz~\cite{hug68} that
"in the many body problem and in statistical mechanics one studies systems with infinitely many degrees of freedom.
Since actual systems are finite but large, it means that one studies a model which not only is mathematically
simpler than the actual system, but also allows a more precise formulation of phenomena such as phase transitions, transport processes, 
which are typical for macroscopic systems. How does one deal with infinitely large systems. The traditional approach 
has been to consider large but finite systems and to take the thermodynamic limit at the end." 
%
%

%
%
%
\section{The Thermodynamic Limit in Statistical Thermodynamics}\label{sec2.2}
%
%
The macroscopic equilibrium thermodynamics~\cite{hon3} can be considered as a limiting case of statistical mechanics. 
This limit was termed by the \emph{thermodynamic limit}. 
The  thermodynamic limit~\cite{petqsm95,bwid02,dorl,dpet,bb,zub74,amun70,acomp89}  or infinite-volume limit gives  the results which are independent of which ensemble 
was employed and independent of size of the box and the boundary conditions at its edge.
Hence the thermodynamic limit is a mathematical technique for modeling  macroscopic systems by 
considering them as infinite composition of particles (molecules). 
The question of existence of these thermodynamical limits  is rather complicated and poses 
lots of mathematical problems~\cite{dorl,leb68,minl,dru99,hug82}. The mathematical theory of thermodynamic limit
is too involved to go into here, but it was discussed thoroughly in 
Refs.~\cite{dorl,leb68,minl,dru99,hug82,mef64,dru63,dru64,fish65,fishru66,dru67,gamir67,penro68,galla70,araki70,lnph73,open78,grif82,geor95}\\
To simplify the problem, sometimes it is convenient to replace the thermodynamic 
limit by working directly with systems defined on classical configuration spaces of infinite volume. 
In this case, one may expects that since these systems tend to show continuous spectra the relevant functions
become relatively well behaved functions.
In a certain case the thermodynamic limit is equivalent to a properly defined
continuum limit~\cite{acomp89}. 

The essence of the continuum limit is that all microscopic fluctuations are suppressed.
The thermodynamic limit excludes  the influence of surface effects. It is defined by~\cite{acomp89}
\begin{equation}\label{3.1}
\lim_{V \rightarrow \infty}
\begin{cases}
V \rightarrow \infty, \\
V/N, E/N   \, \textrm{ constant (microcanonical ensemble)}, \\
V/N,  T   \, \textrm{constant (canonical ensemble)}, \\
 \mu, T  \,  \textrm{constant (grand canonical ensemble)}.  
\end{cases}
\end{equation}   
Thus, in the thermodynamic limit, surface (boundary) effects becomes negligibly small in comparison with the bulk
properties~\cite{petqsm95,bwid02,dorl,dpet,bb,zub74,amun70,acomp89}. 

It is of importance to recall that $N$ and  $V$ are extensive parameters. They are proportional to $V$ when
$V/N = \textrm{const}$. Contrary to this, the parameter $\theta = k_{B}T$ is intensive. It has a finite value as
$V \rightarrow \infty$ when $V/N = \textrm{const}$. In order to describe infinite systems one normalizes extensive 
variables, i.e. those that are homogeneous of degree 
one in the volume, by the volume, keeps fixed the density, i.e. the number of particles per volume, and takes the limit 
for $ N,V$  tending to infinity.  It is at the thermodynamic limit that the additivity property of macroscopic 
extensive variables is obeyed.  \\
The core of the problem lies in establishing the very existence of a thermodynamic limit~\cite{bogol46,bog62,bpk69,petri89,petr09} (such as $N/V = $ const, 
$V \rightarrow \infty$) and its evaluation for the quantities of interest.
Of course, the problem of existence of these thermodynamical limits is extremely complicated mathematical 
problem~\cite{dorl,leb68,minl,dru99,hug82,mef64,dru63,dru64,fish65,fishru66,dru67,gamir67,penro68,galla70,araki70,lnph73,open78,grif82,geor95}  (sometime it 
could be convenient to replace the thermodynamic limit by working directly with systems defined on classical configuration 
spaces of infinite volume, etc.). \\
It was established~\cite{mef64}  that the free energy is the  thermodynamic potential of a system subjected to the 
constraints constant $T,V,N_{i}$. 
To clarify the problem of the thermodynamic limit, let us consider the logarithm of the partition function $Q  (\theta, V, N)$
\begin{equation}\label{3.2}
 F (\theta, V, N) =   - \theta  \ln  Q  (\theta, V, N).
\end{equation}
This expression determines the free energy $F$ of the system on the basis of canonical distribution. The standard way of
reasoning in the equilibrium statistical mechanics do not requires the knowledge of the exact value of the
function $F (\theta, V, N)$. For real system it is sufficient to know the thermodynamic (infinite volume) 
limit~\cite{bogol46,bog62,bog49,bpk69,petri89,hill56,hill60,minl,dru99,mef64,leb68}
\begin{equation}\label{3.3}
 \lim_{N \rightarrow \infty}   \frac{F (\theta, V, N)}{N}\vert_{V/N = \textrm{const}} =  f(\theta, V/N).
\end{equation}
Here $f(\theta, V/N)$ is the free energy per particle. It is clear that this function determines all the thermodynamic 
properties of the system~\cite{dpet,bb,zub74,leb68}.
Thus, the thermodynamic behavior of a system is asymptotically approximated by the results of statistical 
mechanics as $N$ tends to infinity, and calculations using the various ensembles used in statistical mechanics 
converge~\cite{dpet,hill56,hill60,bb,zub74,open78,geor95}.\\
The importance of thermodynamic limit or infinite-volume limit was first mentioned by N.N. Bogoliubov in his seminal 
monograph~\cite{bogol46,bog62}.  
That monograph  describes methods which gave a rigorous mathematical foundation for the limiting
transition in statistical mechanics, using the formalism of the Gibbs' canonical ensemble. A general
formalism was developed for establishing of the limiting distribution functions in the form of formal series in powers
of the density.\\
Later on, in 1949, N.N. Bogoliubov published (with B. I. Khatset) a short article on this 
subject~\cite{bog49} where they formulated briefly their results. 
Here the foundations were developed for a rigorous mathematical description of infinite systems
in statistical mechanics.  These works~\cite{bogol46,bog62,bog49,bpk69,petri89}
gave, in principle,  a full solution to the mathematical problem arising during consideration of the limiting transition
$N \rightarrow \infty$ in systems described by a canonical ensemble, for the case of positive binary particle interaction
potential and sufficiently small density. In this approach the system of equations for the distribution functions
was treated in essence as an operator equation in Banach space. Unfortunately, the methods developed in these papers 
were not known at that time to other investigators in mathematical statistical mechanics.\\
Independently  L. Van Hove~\cite{lvh49,lvh50,legvh} studied the behaviour 
of the statistical system in the limit in which the volume of the system becomes infinitely large.  
He analyzed the problem and found that in the grand ensemble it is only in this limit that phase transitions, in the form 
of mathematically sharp discontinuities, can appear. Thus the thermodynamic limit has reformulated as a pure mathematical 
problem from which certain complications should be  removed. 
The proof of Van Hove contained some mathematical shortcomings and was improved by  
Ruelle~\cite{dru63,dru64,dru67,leb68} and  Fisher~\cite{mef64,fish65,fishru66}.\\
In his paper~\cite{dru63}
Ruelle    suggested a similar to the Bogoliubov-Khatset approach~\cite{bog49} to the study of the systems of equations for 
distribution functions.  He used the formal method of a large canonical ensemble, which simplified
his task in formulating a basis for the limit transition. At the same time Ruelle was able to consider a more wide
class of potential functions  by using the very ingenious idea of making the original equations
for the distribution functions symmetrical. Ruelle  has considered
the well-known Kirkwood-Salsburg equations~\cite{dpet,bras75,swend76,hmor75,hmor77,hmor81}, i.e.
the set of integral equations which form a linear inhomogeneous system for the (generic) distribution functions $f_{A}(x)$. 
In his paper, Ruelle  has taken advantage of the linear structure
of the Kirkwood-Salsburg  equations and has shown how these equations
may be transformed into a single equation for $f_{A}(x)$ in the Banach space. This work has stimulated a series of articles
devoted to studies of the thermodynamic limit in various systems. For example, 
G. Gallavotti and S. Miracle-Sole~\cite{gamir67}  studied   the thermodynamic limit for a classical system of particles
on a lattice and proved the existence of infinite volume correlation functions for a
"large" set of potentials and temperatures.\\ 
The complete mathematical treatment of the thermodynamic limit problem was given by N.N. Bogoliubov and collaborators 
in 1969 in their fundamental paper~\cite{bpk69}. 
That paper formulated  a rigorous mathematical description of the equilibrium state of the infinite system
of particles on the basis of canonical ensemble theory. A proof is given of the existence
and uniqueness of the limiting distribution functions and their analytical dependence on density.
Results have been achieved by using the methods which were based
on the application of the theory of Banach spaces to the study of the equation for the
distribution functions.\\
Bogoliubov and co-authors  showed that
in order to obtain thermodynamic relations on the basis of statistical mechanics one requires to study
systems with an infinite number of degrees of freedom. Such systems are derived from finite systems when
there is an infinite increase in the number of particles $N$ accompanied by a proportional increase in the
volume $V$.  Here difficult problems arose, associated with the rigorous mathematical
basis for the limiting transition as $N \rightarrow \infty$. To solve these problems, as authors showed~\cite{bpk69},   
the formalism of the canonical ensemble supplied with the mechanism of distribution functions is appropriate for the case.\\
They gave  a rigorous mathematical description, based on the theory of the
canonical ensemble, of the equilibrium state (at low density) of infinite systems of particles, whose
interaction potential is free from the restriction of positiveness, and satisfies the Ruelle condition~\cite{dru63}.
Both the methods, i.e., the method  of Bogoliubov-Khatset, and the Ruelle method of symmetrization were used.
For this aim  the relations between the distribution
functions in a finite volume, which for the limit 
transition become the  Kirkwood-Salsburg equations were derived. In contrast with the case of a large canonical ensemble, for a Gibbsian ensemble in a 
finite volume there are generally no equations for the distribution functions: the appropriate equations appear only
after the limit transition to infinite volume. This led to new problems, in comparison with the case of a
large canonical ensemble.
Then   a theorem for the existence and uniqueness of a solution of the Kirkwood-Salsburg equations for the 
potentials satisfying the Ruelle condition was proved. In addition, a clear estimate 
was given for the densities for which the solution is a series of interactions. 
As result of their analysis, a theorem was established concerning the
analytical nature of the dependence of the limit distribution functions on the density.
A proof of the existence of limit distribution functions when the number
of particles in the system tends to infinity was given as well. The uniqueness of these limit functions
was established and proved rigorously.
Thus the paper   by Bogoliubov, Petrina  and Khatset~\cite{bpk69}  and also the classical paper of Bogoliubov and
Khatset~\cite{bog49}  have established the existence of limiting distribution functions for the microcanonical ensemble in
the case of low densities.\\
In the paper by Simyatitskii~\cite{sim71},
some of the arguments and proofs in the paper by Bogoliubov, Petrina  and Khatset~\cite{bpk69} were
simplified. He obtained the same results using essentially the same methods
but by a somewhat shorter path.
The simplifications were achieved by the use of the apparatus of correlation
functions rather than distribution functions. In addition, a more detailed investigation was
made of the question of the equality of the limiting correlation functions of the microcanonical
and grand canonical ensembles in the case of low densities.
In addition, Simyatitskii~\cite{sim71} have been able to avoid many
tedious estimates by referring simply to the results by Dobrushin and Minios~\cite{minl67,minl68}, who
proved an important theorem about   the existence of a limit of the ratios of the microcanonical partition functions.  
On the basis of these results, Simyatitskii also investigated in
detail the question of the equality of the limiting correlation functions of the grand canonical and microcanonical
ensembles for the usual thermodynamic relationship between the density $n$ and the activity $z$ in agreement with the
result by Bogoliubov, Petrina  and Khatset~\cite{bpk69,petri89}.\\  
Kalmykov~\cite{kalm78} analyzed the problem further.
The main aim of his paper was to derive an expression for the thermodynamic potential   in terms
of the limit correlation functions   for classical systems of identical monatomic molecules. 
For single-component systems of hard spheres with binary interaction, the free energy was
expressed in terms of the limit correlation functions of the canonical ensemble. Some
properties of the configuration integral were investigated and estimates obtained for the
correlation functions. His work was based also on the classical results by Bogoliubov, Petrina  and  Khatset~\cite{bpk69,petri89}
and Dobrushin and Minlos~\cite{minl67,minl68}.\\
It is known that in specific physical applications it is important
to have approximate equations for some quantities from which
correlation functions and the equation of state may easily be obtained.
Gonchar and   Rudyk~\cite{gon92}  used this idea  to made a further progress. 
A new set of strict equations for correlation
functions of equilibrium classical statistical mechanics was proposed. The solution was
constructed for the pair repulsive interaction potential at arbitrary values
of activity $z$ and temperature with the help of some nonlinear monotonically
increasing map $L$. In addition,
Gonchar and   Rudyk proved that the radial distribution function oscillates at low density in a
system with a short-range nonnegative potential and investigated the branching
of the solutions of an approximate equation of state.\\
The existence of thermodynamics for real matter with Coulomb forces was
proved by Lieb and Lebowitz~\cite{lile69,lile72}.
They established the existence of the infinite volume (thermodynamic) limit
for the free energy density of a system of charged particles, e.g., electrons and
nuclei. These particles, which are the elementary constituents of macroscopic
matter, interact via Coulomb forces. The long range nature of this interaction
necessitates the use of specific methods for proving the existence of the limit. It was
shown that the limit function has all the convexity (stability) properties required
by macroscopic thermodynamics. They found that for electrically neutral systems, the limit
functions was domain-shape independent, while for systems having a net charge
the thermodynamic free energy density was shape dependent in conformity with
the well-known formula of classical electrostatics. The analysis was based on the
statistical mechanics ensemble formalism of Gibbs and may be either classical
or quantum mechanical. The equivalence of the microcanonical, canonical and
grand canonical ensembles was demonstrated also.\\
H. Moraal~\cite{mor75}  shown, that the configurational partition function for a classical system of molecules interacting with 
nonspherical pair potential is proportional  to the configurational partition function for a system of particles 
interacting with temperature-dependent 
spherical k-body potentials. Therefore, the thermodynamic limit for nonspherical molecules exists if the effective 
k-body interaction is stable and tempered. A number of criteria for the nonspherical potential were developed which 
ensure these properties. In case the nonsphericity is small in a certain sense, stability and temperedness of the 
angle-averaged nonspherical potential are sufficient to ensure thermodynamic behaviour.\\
Heyes and  Rickayzen~\cite{rick07} have investigated in detail a role of
the interaction potential $\Phi (r)$  between molecules (where $r$ is the pair separation). This quantity is the key input 
function of statistical mechanical theories of the liquid state.
They applied the pair interaction stability criteria of Fisher and Ruelle~\cite{fishru66} to establish the range of 
thermodynamic stability for a number of simple analytic potential forms used for condensed matter theory and modelling
in the literature. In this way they identified the ranges of potential parameters where, for a given potential, the system 
is thermodynamically stable, unstable and of uncertain stability. This was further explored by 
carrying out molecular dynamics simulations on the double Gaussian potential in the stable and unstable regimes. 
It was shown that, for example, the widely used exponential-6 and Born-Mayer-Huggins alkali halide potentials produce 
many-particle systems that are thermodynamically unstable. Thus they  have been able to decide the stability or instability 
of potentials which are the difference of two Gaussians or of two exponentials for all real positive values of their 
parameters. The parameter ranges of instability of the generalized separation-shifted Lennard-Jones and 
so-called \emph{SHRAT} potential systems were established in this work.\\
Additional discussion of the applications of the thermodynamic limit in concrete situations were considered by
Styer~\cite{fsty04,fsty04a}. In particular, it was demonstrated that the widely used microcanonical "thin phase space limit" must be 
taken \emph{after} taking the thermodynamic limit.\\
Some important aspects of the nonequilibrium, thermostats, and thermodynamic limits were studied thoroughly 
by Gallavotti and Presutti~\cite{presu10}. They studied many important aspects of the problem, but   left open the
main problem, namely what can be said about the limit $t \rightarrow \infty$, i.e., the study of the stationary states 
reached at infinite time. Instead, a conjecture has been proposed: the limit will be an equilibrium Gibbs distribution
at some intermediate temperature.
%
%
%
\section{Equipartition of Energy}\label{}
%
%
%
In spite of that the problem of equipartition   of energy~\cite{jhj25}  in classical statistical mechanics is an old issue, it is still 
of interest because it can be used to understand better some of the background of statistical mechanics.
The essential problem in statistical thermodynamics is to calculate the distribution of a given amount of energy $E$ 
over $N$ identical systems. The basic statement in statistical mechanics,  which   also known as the equal \emph{a priori} 
probability conjecture,  is the one of the main postulates of the equilibrium statistical 
mechanics~\cite{jhj25,tolm,hill56,hill60}.
The equipartition conjecture rests essentially upon the hypothesis that for any
given   isolated system in equilibrium, it is valid that the system is found with equal probability in each of its 
accessible microstates. The equipartition hypothesis (or theorem) originated in the molecular theory of gases~\cite{jhj25}.
The equipartition theorem states that each degree of freedom contributes $1/2 R T$ to the molar internal energy,
$E$, of a gas. It will be of interest to give here the original Jeans~\cite{jhj25} formulation:
\begin{quote}
\em The energy to be expected for any part of the total energy which can be expressed as a sum of squares 
is at the rate of $1/2 R T$   for every squared term in this part of the energy.
\end{quote}
A gas that consists of individual atoms (like $He$, $Ne$, $Ar$) has a low heat capacity because it has few degrees of
freedom. The atoms can move freely in space in the $x-$, $y-$, or $z$-directions. This translational motion
corresponds to $n = 3$ degrees of freedom. However, atoms have no other types of internal motions such as vibrations or
rotations, so the total number of degrees of freedom for a monatomic system is equal to $3$.
Once the degrees of freedom are determined, the internal energy is calculated from the equipartition theorem,
\begin{equation}\label{4.1}
 E = n (1/2 RT).
\end{equation}
For example, the monatomic gas exhibits only $3$ degrees of freedom. Therefore, the prediction from the equipartition
theorem for the molar internal energy is $ E =  (3/2 RT)$.\\
For diatomic molecules along with linear and nonlinear polyatomic molecules in the gas phase, the number of
degrees of freedom can be determined and therefore the theoretical internal energy and heat capacity can be predicted.
In addition to the 3 translational degrees of freedom, contributions from rotational and vibrational degrees of freedom
must be considered.\\
For diatomic and linear polyatomic molecules, rotational motion contributes $2$ degrees of freedom to the total, while
for nonlinear polyatomic molecules, rotational motion contributes $3$ degrees of freedom. For diatomic and linear
polyatomic molecules, vibrational motion contributes $2(3N-5)$ degrees of freedom to the total, while for nonlinear
polyatomic molecules, vibrational motion contributes $2(3N-6)$ degrees of freedom, where $N$ is the number of atoms
in the molecule. Using these rules, the total number of degrees of freedom can be determined and the
equipartition theorem can then be used to determine a theoretical prediction for the molar internal energy and
the heat capacities.\\
Thus the classical energy equipartition theorem constitutes an important point in equilibrium statistical physics, 
which has been widely discussed and used. \\
In its simplest version  the equipartition principle   deals with the contribution to the \emph{average energy}
of a system in thermal equilibrium at temperature $T$ due to quadratic terms in the Hamiltonian. More precisely, 
it attests that any canonical variable $x$ entering the Hamiltonian through an additive term proportional
to $x^{2}$ has a thermal \emph{mean energy} equal to $ k_{B} T/2$  , where $k_{B}$ is the Boltzmann constant. 
The most familiar example is provided by a three dimensional classical ideal gas.\\
Thus, it should be emphasized that the equipartition principle is a consequence of the quadratic form of terms in the 
Hamiltonian, rather than a general consequence of classical statistical mechanics.
Note, however, that the principle of equipartition is a strictly classical concept, that is, the degree of freedom contributed much  
should be such that $\Delta \varepsilon/k_{B} T $  is small in passing from one level to another.\\
The generalized equipartition principle~\cite{tolm,hill56,hill60,donald00} formulates its essence in the following form.
Let us consider a classical many-particle system of $N$ interacting particles with the Hamiltonian $H (p, q)$. Let $x_{j}$
be one of the $3 N$ momentum components or one of the $3 N$ spatial coordinates. Then the following equality will be hold
\begin{equation}
\label{4.2}  \Bigl \langle x_{i}  \frac{\partial H}{\partial x_{j}}  \Bigr \rangle  = k_{B} T   \delta_{ij}.
\end{equation}
Here $\langle \ldots \rangle$ is the relevant ensemble average. 
It is clear that this equality can only hold asymptotically in the thermodynamic limit.\\
There are more general and advanced formulations~\cite{pland78,phil83,lima00} of the generalized equipartition principle.
Nevertheless, the equipartition, in principle, should be valid in
the thermodynamic limit only. In addition, the equipartition principle yields a direct and intrinsic method for the
definition of the absolute temperature~\cite{tolm,hill56,hill60,donald00,verb83}, irrespective of the interaction or
the phase state. The problem of the consistent definition of the temperature for small systems, such as clusters, etc.,
is under current intensive investigation~\cite{verb83,makar,falci11,letic11}. There are many various applications of the
generalized equipartition principle, for example, application  to the phenomenon of laser cooling and the  equipartition   of energy
in the case of radiation-atom interaction~\cite{menon89}.\\
These are the conclusions arrived at from a study of the equipartition of energy in many-particle systems based on the 
classical dynamics of systems studied.
 Moreover, the presence of the quadratic form of terms in the Hamiltonian was established as decisive. 
Since the mid fifties the intensive studies of the equipartition of energy for nonlinear systems began~\cite{ford61,ford92}. 
Nonlinear effects are of the greatest importance in various fields of science.  
In the last decades a remarkable and fundamental development has occurred in the theory of nonlinear systems,
leading to a deeper understanding of the interrelation of classical and quantum mechanics and statistical 
mechanics~\cite{lgalga85,lgalga92,lgalga93,lgalga04}.\\
The general importance of the nonlinearity for many-particle systems was demonstrated clearly by Ulam, Fermi and Pasta in their seminal 
study~\cite{ford61,ford92}. It was shown that the lack of
equipartition of energy observed by Ulam, Fermi and Pasta for certain nonlinear systems has serious and deep reasons.
Numerous authors have investigated  and explored this fascinating
field~\cite{lgalga85,lgalga92,lgalga93,lgalga04,luka95,luka96,sola97,luka99,dcamp05,lgalga05}, covering much the same
ground of the interrelation of classical and quantum mechanics and statistical mechanics.\\
Galgani~\cite{lgalga93} has presented the point of view of L. Boltzmann on energy equipartition, which is not so well known.
Boltzmann was confronted with the essential qualitative difficulties of classical statistical mechanics of his 
time~\cite{wu75}. The main message is that, according to Boltzmann, the two questions, equipartition and Poincare 
recurrence~\cite{hlf1,hem58,cerc,lav08,bolt09}, "should be treated on the same foot". Roughly speaking, in connection with the problem of equipartition of energy, which seemed to demolish
classical  statistical mechanics, Boltzmann foresaw a solution of the same type he had afforded for the Poincare recurrence
paradox~\cite{hlf1,hem58,cerc,lav08,bolt09}, in the sense that the problem does not occur for finite, "enormously long", times.\\
An averaging theorem for Hamiltonian dynamical systems in the thermodynamic limit was derived by 
A. Carati~\cite{car07} in connection with the the foundations of statistical mechanics.
This theorem helps to understand better some essential
feature of the Fermi-Pasta-Ulam phenomenon: the energy remains confined
to the low frequency modes, while the energies (i.e., up to a factor, the actions)
of the high frequency modes remain frozen up to very large times.
It was shown how to perform some steps of perturbation theory if one
assumes a measure-theoretic point of view, i.e. if one renounces to control
the evolution of the single trajectories, and the attention is restricted to
controlling the evolution of the measure of some meaningful subsets of phase{
space. For a system of coupled rotators, estimates uniform in $N$ for finite
specific energy were obtained in quite a direct way. This was achieved by
making reference not to the $sup$ norm, but rather, following Koopman and
von Neumann, to the much weaker $L^{2}$ norm.\\
Hence, it was established that there are various reasons for lack of the equipartition of energy~\cite{carat07,koz08,carat00,haro00}.    
In this context, it was said~\cite{haro00} that "one of the basic problem of statistical mechanics is to decide its range 
of applicability, in particular, the validity of the equipartition of energy. Deciding what are the boundaries of applicability of statistical
mechanics has become one of the fundamental problems not only for the foundations but, indeed, for the applications".
%
%
%
%
%
\section{Ensemble Equivalence and Nonequivalence}
%
It is well known that the equilibrium thermodynamics~\cite{hill56,hill60,zub74,donald00} of any type of \emph{normal} large 
system (e.g. a monoatomic gas) can be derived using any one of the statistical equilibrium Gibbs ensembles (microcanonical, 
canonical and grand canonical). However, there some subtleties, which should be taken into account properly. To see this point
clearly, it will be useful to remind that, when considering a monoatomic  ideal gas, each of the three ensembles 
will lead to the known equation of state $P V = N k_{B} T$. From the other hand, it is also well known that in canonical
ensemble the number of particles $N$ is fixed, whereas in grand canonical ensemble $N$ is not fixed and can fluctuate. All the
standard considerations~\cite{hill56,hill60,zub74,donald00} of the ensemble equivalence in Gibbs' statistical mechanics are based on the fact that the
fractional fluctuations of $N$ are very small, $\Delta N/N \sim 1/ \sqrt{N}$.\\
The conceptual basis of statistical mechanics and thermodynamics is relatively well established~\cite{isih1,isih2}
and it was shown in various ways~\cite{aboc60,boc63,sork79,fern94,howl03,norm13} that \emph{normal} systems with huge 
degrees of freedom satisfy the laws of statistical mechanics.\\
The question of the ensembles equivalence was considered by various authors. Considerable literature has developed on 
this subject~\cite{open78,geor95,lind66,lind67,lind68,lgalg69,lgalg70,lgalg71,akhal69,rminl72,gura07,adler08}.
A. M. Khalfina~\cite{akhal69} investigated the limiting equivalence of the canonical and grand canonical ensembles
for the low density case.
In that paper it was shown that the limiting Gibbs distribution, whose existence was
established previously by starting from the grand canonical ensemble, can also be
obtained by starting from the canonical ensemble, and both distributions coincide
when a certain relation exists~\cite{rminl72} between the parameters $\beta$ and $\mu$ (for fixed $\beta$).
The proof was based on the local limit theorem for the number of particles.\\
It was shown  by Adler and Horwitz~\cite{adler08}, that complex quantum field theory 
can emerged as a statistical approximation to an underlying generalized quantum dynamics. 
Their approach was based on the already established formalism  of application of statistical mechanical methods 
to determination of the canonical ensemble governing the equilibrium distribution of operator initial values. 
Their result was obtained by the arguments based on a Ward identity 
(analogous to the equipartition theorem of classical statistical mechanics). Adler and Horwitz~\cite{adler08} constructed 
in their work a microcanonical ensemble which 
forms the basis of this canonical ensemble. That construction enabled to them to define the microcanonical 
entropy and free energy of the field configuration of the equilibrium distribution and to study the stability 
of the canonical ensemble. They also studied the algebraic structure of the conserved generators from which the 
microcanonical and canonical ensembles were constructed, and the flows they induce on the phase space. \\
Although the ensemble equivalence holds for \emph{normal} large system, we will mention,
mainly by reference only, a few examples of systems where 
the nonequivalence of Gibbs ensembles occur~\cite{touch05,touch06,touch09,ninn12}
by various reasons.\\
Some objection to the standard arguments of the ensembles equivalence were put forward 
recently~\cite{touch05,touch06,touch09}. According to this point of view
some researchers have found examples of statistical
mechanical models characterized at equilibrium
by microcanonical properties which have no equivalent
within the framework of the canonical ensemble. The
nonequivalence of the two ensembles has been observed
for these special models both at the thermodynamic and the macrostate
levels of description of statistical mechanics of these systems. This is a contradiction with
J. W. Gibbs~\cite{gibbs},   who insisted that the canonical ensemble should be equivalent to the microcanonical
ensemble in the thermodynamic limit. In this limit,
\emph{the thermodynamic limit}, the system \textbf{should} thus appear
to observation as having a definite value of energy - the very \emph{conjecture} which the microcanonical ensemble is
based on. The conclusion then apparently follows, namely: both the microcanonical and the canonical ensembles
should predict the same equilibrium properties of many-body systems in the thermodynamic limit of these
systems independently of their nature. The fluctuations of the system's energy should become negligible
in comparison with its total energy in the limit where the volume of the system tends to infinity.\\ 
H. Touchette and co-authors~\cite{touch05,touch06,touch09} attempted
to give relevant physical interpretation and an accessible explanation of the phenomenon of
nonequivalent ensembles.\\
In particular, H. Touchette and co-authors~\cite{touch05,touch06,touch09} investigated various aspects of
generalized canonical ensembles and corresponding ensemble equivalence.
They introduced a generalized canonical ensemble obtained by multiplying the usual Boltzmann weight factor
$\exp (- \beta H)$   of the
canonical ensemble with an exponential factor involving a continuous function $g$ of the Hamiltonian $H$. They
focused   on a number of physical rather than mathematical aspects of the generalized canonical ensemble. 
The main result obtained is that, for suitable
choices of $g$, the generalized canonical ensemble reproduces, in the thermodynamic limit, all the microcanonical
equilibrium properties of the many-body system represented by $H$ even if this system has a nonconcave
microcanonical entropy function. This is something that in general the standard $(g = 0)$ canonical ensemble
cannot achieve. Thus a virtue of the generalized canonical ensemble is that it can often be made equivalent to
the microcanonical ensemble in cases in which the canonical ensemble cannot. The case of quadratic $g$
functions was discussed in detail; it leads to the so-called Gaussian ensemble.\\
Very recently it was pointed by G. De Ninno  and D. Fanelli~\cite{ninn12}, that
classical statistical mechanics most commonly deals with large systems, in which the interaction range among components 
is much smaller than the system size. In such "short-range" systems, energy is normally additive and statistical 
ensembles are equivalent. The situation may be radically different when the interaction potential decays so slowly 
that the force experienced by any system element is dominated by the interaction with far-away components. In 
these "long-range" interacting systems  energy is not additive. Well-known examples of non-additive "long-range" interacting
systems are, for instance, found in cosmology (self-gravitating systems) and plasma physics applications, 
where Coulomb interactions are at play. The lack of additivity, together with a possible break of ergodicity, may 
be at the origin of a number of peculiar thermodynamic behaviours: the specific 
heat can be negative in the microcanonical ensemble, and temperature jumps may appear at microcanonical 
first-order phase transitions. When this occurs, experiments realized on isolated systems give a different result 
from similar experiments performed on systems in contact with a thermal bath. As a consequence, the canonical and 
microcanonical statistical ensembles of long-range interacting systems may be non-equivalent.\\
G. De Ninno  and D. Fanelli~\cite{ninn12}, discussed out-of-equilibrium statistical ensemble nonequivalence.
They considered a paradigmatic model describing the one-dimensional motion of N rotators coupled through a mean-field 
interaction, and subject to the perturbation of an external magnetic field. The latter was shown to significantly alter 
the system behaviour, driving the emergence of ensemble nonequivalence in the \emph{out-of-equilibrium phase}, as signalled 
by a negative (microcanonical) magnetic susceptibility. The thermodynamic of the system 
was analytically discussed, building on a maximum-entropy scheme justified from first principles. Simulations confirmed 
the adequacy of the theoretical picture. Ensemble nonequivalence was shown to rely on a peculiar phenomenon, different 
from the one observed in previous works. As a result, the existence of a convex intruder in the entropy was found to be a 
necessary but not sufficient condition for nonequivalence to be (macroscopically) observed. Negative-temperature states 
were also found to occur. These intriguing phenomena reflect the non-Boltzmanian nature of the scrutinized 
problem and, as such, bear traits of universality that embrace equilibrium as well as out-of-equilibrium regimes.\\
However, it should be emphasized that this field of researches is still
under debates and the thorough additional investigations in this direction should be 
carried out~\cite{smith11,calle11,petr,leeh02}.

%
%
%
\section{Phase Transitions}
%
%
%
The aim of statistical mechanics is to derive 
the properties of macroscopic systems from the properties of the individual particles and their interactions. 
In particular it is the task of statistical mechanics to give an explanation of phase transitions, 
transport phenomena and the approach to equilibrium in the course of time for a non-equilibrium system.
Physicist have been trying to understand the occurrence of phase transitions by use of statistical mechanics
since the famous dissertation of van der Waals in 1873.\\
The problem of phase transitions in the interacting many-particle systems has been studied intensively
during the last decades from both the experimental and theoretical viewpoints~\cite{minl,dobr94,ptcp20,sinai82,geor88,domb,herb07,sole11}.
Phase transitions occur in both, equilibrium and nonequilibrium systems.
Typical examples of the equilibrium phase transitions are the transitions
between different states of matter (solid, liquid, gaseous, etc.) or the transition from
normal conductivity to superconductivity.\\
In the vicinity of a phase transition point~\cite{domb,herb07},
a small change in some external control parameter (like pressure or temperature)
results in a dramatic change of certain physical properties (like specific heat or electric
resistance) of the system under consideration.
Many  aspects of the theory of phase transitions are related in one way or another with the thermodynamic limit
transition procedure~\cite{minl,dru99}. This is rather evident from the fact that
an equilibrium phase transition is defined as a nonanalyticity of the free-energy density $F/N$.\\
Phase transitions have been an important part of statistical mechanics for many years.
During the last decades the mathematical theory of the phase transitions~\cite{minl,dobr94,dobr,ptcp20,sinai82,geor88}  
achieved a marked progress, in particular in a systematic study of the (quantum) mechanics of systems with infinitely many degrees of freedom.
The theory of operator algebras, in particular $C^{*}$-algebras~\cite{hug68,hug72}, plays an important part in 
these developments.\\ 
Although, in certain models, one can prove the existence of a phase transition,  for instance in the 
Ising model in two and more dimensions with zero external field~\cite{minl,dobr94,ptcp20,sinai82,geor88,domb,herb07,sole11}, theoretically the situation 
with respect to phase transitions in general still is not fully understood. 
More recently phase transitions have become an object of intensive studies in computer science (study of 3-satisfiability), 
combinatorics (birth of the giant component for various random graph models) 
and probability theory (cutoff phenomena for Markov chains).\\
Here we touch briefly of some issues only from the physical viewpoint. Such a physical viewpoint on the essence of the
phase transitions was formulated recently by M. E. Fisher and C. Radin~\cite{fish06}. We shall follow  to that work  
reasonably close because of its remarkable transparency and clarity.\\ 
According to M. E. Fisher and C. Radin~\cite{fish06},
there are various thermodynamic variables one can use to describe matter in
thermal equilibrium, some of the common ones being: mass or number density $N/V$,
energy density $E/N$, temperature $T$, pressure $P$, and chemical potential $\mu$.  
By definition the states of a "simple" system can be parameterized
by two such (independent) variables, in which case the others can be regarded as
functions of these. We will assume we are modelling a simple material. Then a
particularly good choice for independent variables is $T$ and $\mu$.\\  
M. E. Fisher and C. Radin~\cite{fish06} remarked that
it is a fundamental fact of thermodynamics that the pressure $P$ is a convex function of these variables,
and, in particular, this convexity embodies certain mechanical and thermal stability
properties of the system. Moreover, all thermodynamic properties of the material
can be obtained from $P$ as a function of $T$ and $\mu$ by differentiation.\\
It is worth reminding that   the question of a convexity
of thermodynamic variables was investigated in detail by L. Galgani and A. Scotti~\cite{lgal68,ascot69,galg70}.
They  considered  the usual basic postulate of increase of entropy for an isolated system. In addition,
it was pointed out that that postulate can be formalized mathematically as a superadditivity property of
entropy. This fact has two kinds of implications. It allows one to deduce in a very direct  and 
mathematically clear way stability properties such as $c_{V} \geq 0$ and $K_{T} \geq 0$. 
Here $c_{V} =  \left (  T \partial \mathcal{S}/ \partial T \right )_{V}$ is
a specific heat  and $K_{T} =  - 1/V \left (\partial P/ \partial V \right )_{T}$; the entropy $S$ was
defined through the functional relation $S = \mathcal{S} (E, V, N)$. \\
On this basis L. Galgani and A. Scotti~\cite{lgal68,ascot69,galg70} were able to justify of the equivalence
of various thermodynamic schemes as expressed for example by the fact that the \emph{minimum} property
of the free energy is a consequence of the \emph{maximum} property of entropy.\\ 
The following definitions given below were straightforwardly adapted from Ref.~\cite{fish06}\\
A thermodynamic phase of a simple material is an open, connected
region in the space of thermodynamic states parametrized by the variables $T$ and
the pressure $P$ being analytic in $T $ and $\mu$. Specifically, $P$ is analytic in $T$ and $\mu$,
at $(T_{0}, \mu_{0})$ if it has a convergent power series expansion in a ball about $(T_{0}, \mu_{0})$ that
gives its values. Phase transitions occur on crossing a phase boundary.  
The graph of $P = P(T, \mu)$ is not only convex but (for all reasonable physical
systems) also has no (flat) facets. M. E. Fisher and C. Radin~\cite{fish06} used this fact in their definition 
of phase; without this
property there would typically be open regions of states representing the coexistence
of distinct phases. The essential point is   the choice of independent variables, which
can lead to the appearance of domains representing two or more coexisting phases.
They noted also that in particular the isothermal (i.e., constant $T$) "tie lines" connecting the distinct
phases that can coexist at the range of overall intermediate densities spanned at a
fixed temperature.\\
On their phase diagram~\cite{domb,herb07,sole11,fish06} an intrinsic difference between vapor and liquid
"phases", which can be analytically connected, and between these regions of the
fluid phase and the solid phase, which cannot be so connected may be clearly seen.\\  
M. E. Fisher and C. Radin~\cite{fish06} mentioned that
in the modern literature  an important distinction is made
between "field" variables and "density" variables, which helps to explain various
consequences of the choice of independent and dependent variables.
The foregoing constitutes a "thermodynamic" description of phases and phase
transitions. There is a deeper description, that of statistical mechanics, deeper in
that it allows natural ("molecular") models from which one can in principle compute
the pressure as a function of $T$ and $\mu$.\\ 
In the statistical mechanical description the thermodynamic states are realized
or represented as probability measures on a certain space and the measures still
parameterized by thermodynamic variables, e.g. the two variables,  
specifically temperature $T$ and chemical potential $\mu$).
M. E. Fisher and C. Radin~\cite{fish06}   considered  first a finite system of $N$ particles contained
in a reasonably shaped domain, say $\Lambda$ of volume $V$. In this case the probability
densities in the phase space  $(\mathbf x, \mathbf p)$  for 
particles, will be proportional to the weights $f_{N} (T,\mu,  \mathbf x, \mathbf p)$.\\
The structure of the energy $E_{N}$ is determined only when one settles on the
type of "interactions" the constituent particles can undergo; that not only depends
on the material being modelled but also on what environment. Then they considered
the \emph{ grand canonical}  pressure of the finite-volume system, which  is given
by $P_{V} (T, \mu)$. For reasonable interaction potentials $\Phi$  the pressure $P_{V}$ as
a function of $T$ and $\mu$ will be everywhere analytic.
In order to model a sharp phase transition they considered the thermodynamic limit 
\begin{equation}
\label{ }  P (T, \mu) = \lim_{V \rightarrow \infty} P_{V} (T, \mu).
\end{equation}
Then $P (T, \mu)$ may be identified as the thermodynamic pressure to which the above definitions of a phase 
and a phase transition applies. The proof of the existence of the thermodynamic limit requires certain conditions on
the interaction potential.\\
In the present context, this very clear but terse formulation of the role of the thermodynamic limit  requires
an additional comment.
First, a a few general remarks will be useful.
It is known~\cite{kast10} that to discuss a certain phase transition of interest with the above definition, the free energy
density has to be considered as a function of the relevant control parameters,
i. e. those which, upon variation, give rise to the phase transition.\\
The number of independent intensive variables, $r$, which determine the state of a  heterogeneous system is 
given by the Gibbs phase rule~\cite{hon3},
$$ r = c - \phi + 2,$$ where $c$ is the number of independent components  and $\phi$ is the number 
of phases in the system.\\
For the phase transitions
between the aggregate states of, say, water, the (Gibbs) free-energy density as
a function of temperature and pressure is a suitable choice.  
For spin systems there are at most two such relevant control
parameters, the temperature $T$ and an external magnetic field $H_{\textrm{ext}}$, and therefore  
 the free-energy density $f(T, H_{\textrm{ext}})$ will be a function of the inverse temperature
$\beta = 1/(k_{B}T)$   and the magnetic field $H_{\textrm{ext}}$. 
Quantities like the specific heat or caloric curves which are typically measured in
an experiment are then given in terms of derivatives of the free-energy density. Nonanalyticities
of derivatives may hence lead to discontinuities or divergences in these quantities,
which are experimental hallmarks of phase transitions.\\
Our special interest will be in the emphasizing of the main difficulty in the theory of phase transition
in the many-particle interacting systems. This is the task of the evaluation of partition  functions
associated with particular physical systems of interest. In this context it will be of instruction to discuss 
the concept of the isothermal-isobaric (or $T-P$) ensemble~\cite{hill56,hill60,iked67,kama67}, which is used
in the condensation theory~\cite{hill56,hill60,lew56}.\\
A system (consisting of $N$ molecules) in the isothermal-isobaric ensemble of temperature $T$ and pressure $P$
is described by means of partition function~\cite{hill56,hill60,iked67,kama67}
\begin{equation}
\label{ }  R_{N} (P, T) = \int^{\infty}_{0} d V \sum_{i} \omega_{i}   \exp \left( \frac{- P V - E_{i}}{k_{B} T} \right).
\end{equation}
The equation of state for the imperfect gas   was deduced~\cite{iked67,kama67} in terms of the cluster concept.
Then the properties of imperfect gases and the condensation phenomena were investigated and described in the limit
$N \rightarrow \infty$, employing the concepts of "small",  "large", and  "huge" clusters. What is remarkable,
when authors in their theory~\cite{iked67} have neglected the volume dependence of the cluster integral the obtained 
an unrealistic result: the lower limit of the range of fluctuation in $v = V/N$ has become zero. When, however,
they introduced~\cite{kama67} the volume dependence of the cluster integrals, this lower limit becomes a certain positive value,
corresponding to the volume of the pure liquid. As vas stressed above, phase transitions of a physical systems stem from the
singularities of a \textbf{limiting functions} related to the partition  functions of the system. The 
limit $\langle v \rangle_{\infty}$ ( for $N \rightarrow \infty$) of the ensemble average $\langle v \rangle$
of the specific volume $v = V/N$, which fluctuates in the $(T-P)$  ensemble, was calculated in the form~\cite{iked67,kama67}
\begin{equation}
\label{ }  \langle v \rangle_{\infty} =   \lim_{N \rightarrow \infty} \langle v \rangle = - \lim_{N \rightarrow \infty}
\frac{k_{B} T}{N} \Bigl ( \frac{\partial}{\partial P} \ln R_{N} (P, T)  \Bigr )_{T} = k_{B} T 
\Bigl ( \frac{\partial \ln z}{\partial P}  \Bigr )_{T},
\end{equation}
where $z$ is the activity.\\
This example shows clearly that the procedure of taking the thermodynamic limit  requires very careful performance.
%
%
\section{Small and Non-Standard Systems}
%
%
%
Statistical physics  derives observable (or emergent) properties of
macroscopic matter from the atomic structure and the microscopic dynamics.  
Those characteristics are temperature,
pressure, mean flows, dielectric and magnetic constants, etc., which are essentially
determined by the interaction of many particles (atoms or molecules). The central point of statistical physics
is the introduction of probabilities into physics and connecting them with the fundamental
physical quantity entropy. A special task of this theory was to connect microscopic behavior with 
thermodynamics.\\
From the brief sketch of the statistical thermodynamics, already given above, it should be clear that the 
"normal" thermodynamic systems must be large enough to avoid the influence of the boundary effects.
In statistical mechanics~\cite{haag} one studies large systems and the aim is to derive the macroscopic, or 
thermodynamical properties of such systems from the equation of motion of the individual particles. Due to their large
size, such systems have features such as phase transitions~\cite{minl,dobr94,ptcp20,sinai82,geor88,domb,herb07,sole11,lvh57},
transport phenomena~\cite{zub74,kuz07}, which are absent in small systems. To exhibit such features in full measure
one has to consider the limiting case of infinitely large systems, i.e., systems with infinitely many degrees of
freedom. This means that one has to consider large, but finite, systems and take the thermodynamic limit at the end.\\
However, small systems~\cite{hill02,feshb88,gros01,regu03,busta05,behri05,naud05,gross06} are becoming 
increasingly interesting from both the scientific and applied viewpoints.
Small systems are those in which the energy exchanged with the environment is a few times
$k_{B} T $ and energy fluctuations are observable.  
For example, nanoscience~\cite{emil06,rod06,hnano10a} demands a progressive reduction in
the size of the systems, and the fabrication of the new materials
requires an accurate control over condensation and crystallization~\cite{regu03}.
Small systems found throughout physics, chemistry, and
biology manifest striking properties as a result of their
tiny dimensions~\cite{busta05}. Examples of such systems include magnetic
domains in ferromagnets, which are typically smaller
than 300 nm, quantum dots and biological molecular machines
that range in size from 2 to 100 nm, and solid-like
clusters that are important in the relaxation of glassy systems
and whose dimensions are a few nanometers. There are a big interest  in understanding the properties
of such small systems~\cite{hill02,feshb88,gros01,regu03,busta05,behri05,naud05,gross06}.\\
There are a lot of specificities in describing such systems~\cite{makar,falci11,letic11,leeh02,behri05,naud05,gross06}.
For examples, J. Naudts~\cite{naud05} showed by    slight  modification of the Boltzmann's entropy that it is possible 
to make it suitable for discussing phase transitions in finite systems. As an example, it was shown that the pendulum undergoes a
second-order phase transition when passing from a vibrational to a rotating state.\\
There is an  interest in phase transitions in pores and in the so-called melting of small clusters~\cite{regu03}.  
Although these clusters are equilibrated in a heat bath before being
isolated, when they are isolated each cluster corresponds to a
microcanonical ensemble in which a ''\emph{microcanonical  temperature}''(c.f. Ref.~\cite{feshb88}) must be defined via reference to 
entropy~\cite{regu03}. The act of ''melting'' then becomes a matter of definition, etc.\\
These topics form the new branch of thermodynamics~\cite{naud11}, the so-called   nanothermodynamics and non-extensive thermodynamics. 
They are used to study of those physical systems that have not the property of 
extensivity and are characterized by a small size.
%
%
\section{Concluding Remarks}
%
%
This review was limited to selected topics of the statistical mechanics. The emphasis was on   
the thermodynamic limit, equipartition of energy and equivalence and nonequivalence of ensembles.\\
The analysis carried out in the previous sections shows that from the statistical mechanics point of view,
a thermodynamic system is one whose size is large enough so that fluctuations are negligible. This was shown
very clearly by many authors, e. g., by T. L. Hill~\cite{hill56,hill60} and D. N. Zubarev~\cite{zub74} in their books on 
statistical mechanics and thermodynamics~\cite{hill56,hill60}.
This is a conclusion arrived at from the present study of the problem of the thermodynamic limit.\\
To sum up, the statistical mechanics  is best applied to large systems. Formally, its results are exact only for infinitely 
large systems in the thermodynamic limit.  However, even at the thermodynamic limit, there are still small detectable 
fluctuations in physical quantities, but this has a negligible effect on most sensible properties of a system.
The thermodynamic functions calculated in statistical mechanics should be independent of the ensemble
used in the calculation. But as to the fluctuations, the situation is different. For each environment, i.e. for each ensemble
the problem is different. Moreover, the variable which fluctuate are different~\cite{zub74}.\\
It is the hope of the author that the present short  review  will serve, nevertheless, as a quick
introduction to the subject and will help reader to appreciate vividly a beauty and elegance of statistical mechanics 
as an actual and developing  branch of contemporary science.
%
\section*{Acknowledgements}
%
%
The author recollects with gratefulness discussions
of some of this review topics with N. N. Bogoliubov and D. N. Zubarev. He is also grateful to Luigi Galgani
for numerous stimulating discussions and to Robert Minlos for valuable conversations.
%
%
%

%
%


%

\end{document}